\documentstyle[12pt,aasms4]{article}
\begin{document}

\title{The Infrared Nucleus of the Wolf-Rayet Galaxy Henize 2-10\altaffilmark{1}}
\author{S. C. Beck\altaffilmark{2}}
\affil{School of Physics and Astronomy of the Sackler Faculty of Exact
Sciences and Wise Observatory; Tel Aviv University, 69978 Ramat Aviv, Israel}
\and
\author{D. M. Kelly\altaffilmark{2,3}, J. H. Lacy}
\affil{Department of Astronomy/McDonald Observatory, University of
Texas, Austin, TX 78712}
\altaffiltext{1}{Wise Observatory Preprint 96/77}
\altaffiltext{2}{Guest Observers, Infrared Telescope Facility, which 
is operated by the University of Hawaii under contract through NASA.}
\altaffiltext{3}{Currently at Wyoming Infrared Observatory, University of
Wyoming, Laramie, WY 82071-3905}

\begin{abstract}

We have obtained near-infrared images and mid-infrared spectra of the 
starburst core of the dwarf Wolf-Rayet galaxy He~2-10. We find that the 
infrared continuum and emission lines are concentrated in a flattened 
ellipse 3-4$^{\prime\prime}$ or 150 pc across which may show where a 
recent accretion event has triggered intense star formation.  The 
ionizing radiation from this cluster has an effective temperature of
40,000 K, corresponding to $30M_\odot$ stars, and the starburst is 
$0.5-1.5 \times 10^7$ years old.  

\end{abstract}

\keywords{Galaxies: Individual (He~2-10) --- Galaxies: Starburst ---
Infrared: Galaxies}

\section{Introduction and Observations}

He~2-10 is a blue compact dwarf galaxy and the first galaxy in which
Wolf-Rayet emission features were seen (\markcite{a76}Allen, Wright, 
\& Goss 1976).  It contains a bright star cluster A and a fainter cluster 
B 380 pc (8$^{\prime\prime}$ at the assumed distance of 9.2Mpc) to the 
east of A.  The outer isophotes of the galaxy are smoothly elliptical. 
The kinematics of the molecular and atomic gas (\markcite{k95}Kobulnicky 
et al.~1995) show that He~2-10 is probably the moderately advanced merger 
of two dwarf galaxies.  Optical spectroscopy (\markcite{v92}Vacca \& 
Conti 1992) found hundreds of WR and thousands of O stars in the brighter 
cluster A, and HST UV images (\markcite{c94}Conti \& Vacca 1994) show that 
the hot stars are grouped into knots the size and mass of globular clusters 
but from 1 to 10 Myr old. He~2-10 is thus a ``starburst dwarf''.  Starburst 
dwarfs differ in important ways from other starburst galaxies: they are 
objects whose non-burst star formation rate is almost nil, they have low 
metallicity compared with larger galaxies, and their star formation is not 
driven by the dynamical mechanisms that dominate in spiral galaxies. 

We set out to study the current stellar population of the starburst 
in He~2-10, with the hope of deducing the history of star formation 
activity in this galaxy and finding what may have caused it.  It is known
(\markcite{k89}Kawara, Nishida, \& Phillips 1989) that the brightest region 
of He~2-10 is actually heavily obscured.  We therefore observed this 
galaxy in the infrared where the extinction will be relatively small. 
We obtained spectra of the important infrared emission lines in the 
$8-13\mu$m region with the Irshell spectrograph (described in \markcite{l89}Lacy 
et al.~1989) at the NASA IRTF in May 1995. A wavelength region 
equivalent to 950 km s$^{-1}$ was scanned around each of the [Ne~II] 
$12.8\mu$m, [Ar~III] $8.99\mu$m, and [S~IV] $10.5\mu$m lines. The 
spectral resolution was about 30 km s$^{-1}$. The slit was 
2$^{\prime\prime}$ wide and sampled by 11 pixels along its 
11$^{\prime\prime}$ length.  It was oriented NS for the [Ne~II] 
measurements and EW for the others.  The telescope drifted by as much 
as 1-2$^{\prime\prime}$ during a ten minute observation. The weather was good 
for the [Ne~II] and [Ar~III] observations but unstable for the [S~IV], 
so we will use a [S~IV] measurement with the same instrument from ESO 
made available to us by J.M.~van der Hulst.  The infrared line fluxes 
and details of the observations are in Table I and a typical [Ne~II] 
spectrum is shown in Figure 1. We used a card at ambient temperature and
standard stars for flux calibration and the results are accurate to 20\%.
The data were reduced and analysed with the SNOOPY package 
(\markcite{a92}Achtermann 1992).  

 We also present in Figure 2 J, H and K band continuum images obtained 
for us as a service observation by Mr.~Charles Kaminski using the 
NSFCAM on the NASA IRTF. The spatial scale on these figures is 
0.3$^{\prime\prime}$ per pixel, and the FWHM of stars in the field is
0.9$^{\prime\prime}$. The images were reduced and analysed with the VISTA 
package. Each image in figure 2 is the sum of two or three exposures
in that band which were flatfielded with dome flats. The exposure times 
were different in each band and totalled 6 sec at J, 48 sec at H, and 60 
sec at K; the galaxy was at least 35000 counts above the sky background 
even in these short exposures.  We have not attempted to flux-calibrate the
images. 

\section{Spatial Extent and Structure of the Starburst}

\subsection{The Infrared Continuum}

Optical observations of He~2-10 show that continuum and line emission 
are strongly concentrated in a small clump (Vacca \& Conti's A) in 
the center of the main body of the galaxy.  Corbin, Korista, \& Vacca's 
(\markcite{c93}1993) images show that the central source in the inner 
10$^{\prime\prime}$ extends towards PA 127 (in degrees east of north) 
in $H\alpha$ and in the V band.  The V-band image of the entire galaxy 
has a halo at PA 168, which Corbin et al.~believe is the major axis of 
the object. At $10\mu$m Sauvage, Lagage, \& Thuan (\markcite{s95}1995) 
found an HII region NW of the main clump which they identify with the 
$H\alpha$ extension in that direction.  The near-infrared images in J,H, 
and K bands are less affected by extinction than is the optical and have 
the further advantage that they are good bands for locating giant and 
supergiant stars and dust-obscured O and B stars. The J,H, and K images 
in Figure 2 look very much alike; they each show an elongated ellipse of 
dimension (defined from the contour which is 10\% of peak above the 
background) about 5x3 arcsec extended in PA 130, which we will hereafter 
call "the disk" (because it looks like a disk, not because there is any 
data on its rotation).  If the brightest point of the disk is the center 
of the galaxy then the disk is not symmetric around the center but extends 
further to the NW than to the SE, and on the NW side of the center has a 
slightly less acute (to NS) angle.  We believe that both these effects are 
due to unresolved subsources, notably the $10\mu$m H~II region, probably 
the small clusters of hot stars seen in the UV images, but possibly others 
which are still unknown, rather than to a bar structure. The logs of the 
ratios of J/K, J/H and K/H, which are equivalent modulo a constant to the 
relative J-K, J-H and K-H colors, are in Figure 3. There are no obvious 
structures that could be attributed to dust lanes. (It should be noted that 
the infrared aperture photometry of Johansson (\markcite{j87}1987) is on a 
much larger scale; the entire disk lies inside the smallest aperture used there.) 

We see that all the tracers of active star formation are concentrated in 
the NIR disk. In addition to the near and middle infrared emission and 
the $H\alpha$ described above, the UV star clusters found by Conti \& 
Vacca (\markcite{c94}1994) appear to lie across the disk in a bent line 
(bent in the same sense as the shift across the center seen in the NIR). 
We suggest that the UV star clusters lie on the near, less-heavily 
obscured side of the disk. 

How does the NIR disk, the locus of current star formation, relate to the 
overall structure of the galaxy and its merger history? The accreted CO 
has a velocity structure reminiscent of a galaxy disk with the maximum 
velocity gradient in PA 130, agreeing with the NIR disk. The brightest CO 
is at PA 150, much closer to the PA 160 angle of the galactic halo 
(\markcite{k95}Kobulnicky et al.~1995).  We suggest that the NIR disk is 
the original disk of the galaxy and that accreted CO falling onto it is 
fueling the burst of star formation seen in the UV and IR.  That the 
galactic halo is at a different angle than the disk may record an offset 
between disk and halo that predates the accretion event. It is also 
possible that the accretion process itself, which is well-advanced and 
probably at least $10^8$ years old, distorted the angle of the galactic 
halo.  This form of merger has been little studied so this suggestion 
must remain tentative. 

It should be noted that the 3.6 cm radio continuum observations of 
Kobulnicky (\markcite{k96}1996), the CO observations of Kobulnicky et 
al.~(\markcite{k95}1995), and the H$_2$ measurements of Baas, Israel, 
\& Koorneef (\markcite{b94}1994) do not agree with the infrared continuum 
and emission lines.  The peak of CO emission is 2$^{\prime\prime}$ 
ENE of A, as is a strong source of 3.6 cm radio emission, and the peak of 
shocked H$_2$ is offset 1-3$^{\prime\prime}$ east from A.  It is probable 
that the CO marks a body of gas not currently forming stars and that the 
H$_2$ is shock excited by winds from the starburst impacting the molecular 
cloud. It is harder to explain the offset between the 3.6 cm radio continuum 
and the starburst. That region of the galaxy may be so blanketed by 
extinction that even the 2 and 10 $\mu$m continua and $12.8\mu$m [Ne II] 
emission cannot penetrate; this would be consistent with the concentration 
of molecular gas in that area. Or, the 3.6 cm emission may be from 
supernovae and trace an older stage of the starburst which does not 
currently have detectable infrared continuum or line emission, in which 
case it should be strong at 20cm (although measurements at 20cm with the 
spatial resolution needed to separate the 3.6 cm emission peak from A will 
be hard to achieve). 

\subsection{The [Ne~II] Lines}

We attempted to derive the size of the [Ne~II] emission line source
by comparing the Irshell data on [Ne~II], the strongest line observed,
to stellar profiles. The PSF for the run in which these data were 
obtained was somewhat elliptical due to an optical misalignment, 
which limits the accuracy of this method, but we can say that the 
[Ne~II] emission on peak A is 3$^{\prime\prime}$ in diameter in both 
dimensions (with a probable error of 0.5$^{\prime\prime}$). The spatial 
resolution of the [Ne~II] is enough lower than the infrared and UV images 
that the disk structure seen in the infrared and the small star clusters 
seen in the UV could not be distinguished even if present.  We will therefore
refer to the emission source detected in [Ne~II] as the star cluster. This
will provide consistency with the accepted use of cluster A and B for the 
bright optical sources. But we think it possible and even likely that it will 
break up into smaller and more structured sources under higher resolution. 

The secondary $10\mu$m H~II region was not mapped in [Ne~II].  We looked 
for [Ne~II] on B and did not detect it, which is consistent with the 
relative strengths of the two clusters in other wavelengths and is further 
argument that star formation in cluster B, while it exists, is at too low 
a level to be called a starburst.

If the kinematics of the [Ne~II] are like those of the CO and HI, the FWHM 
would be 50-80 km s$^{-1}$.  Our lines were broadened by a misalignment 
of the dewar optics, but they showed similar profiles to the [Ne~II] 
observed on the same night in the Galactic H~II region W33. The lines in 
W33 are expected to be no more than 50 km s$^{-1}$ wide, so we place an 
upper limit on the lines in He 2-10 of FWHM $\le$ 100 km s$^{-1}$.

The common practise of quoting ``the'' extinction to a galaxy is 
misleading: galaxies and the dust within them can have complex 
three-dimensional structures.  It is somewhat more precise to discuss 
the extinction to a given component of a galaxy.  Vacca \& Conti's 
(\markcite{v92}1992) optical spectroscopy of He~2-10 found $E_{(B-V)}=0.56$ 
or $A_v=1.7$ mag to region A.   Such a high value for $A_v$ implies
that the optical observations could not see through the source.  The 
infrared Brackett line measurements of Kawara et al.~(\markcite{k89}1989), 
which will probe deeper into the source, found $A_v$ of about 17 mag, 
which is consistent with the depth of the $10\mu$m silicate feature 
and the molecular column observed by Kobulnicky et al.~(\markcite{k95}1995). 
He~2-10 is therefore like NGC~5253 (\markcite{b96}Beck et al.~1996) 
and many other galaxies where optical and UV observations see the near 
side of the star forming region and longer wavelength measurements see 
deeper in. The extinction in the infrared, while lower than in the 
optical, is not negligible. Unfortunately the mid-infrared extinction 
law is not well known and seems to differ in different environments 
(\markcite{d89}Draine 1989); furthermore, the dependance of extinction 
on the metallicity is unknown.  Extinction laws in the literature give 
results ranging from 0.6 mag at [Ne~II] and about 1.7 mag at [Ar~III] 
and [S~IV] (\markcite{b78}Becklin et al.~1978) to 0.22 mag at [Ne~II] 
and 0.55 mag at [Ar~III] (\markcite{r84}Roche \& Aitken 1984) for 
$A_v = 17$ mag.  We will discuss the importance of this uncertainty 
below. 

\section{Stellar Population and Star Burst History}

\subsection{Stellar Population}

The O star population of the obscured region can be found from the 
mid-infrared and Brackett lines.  First, we compared the ratios of 
the [Ne~II], [Ar~III] and [S~IV] line fluxes to the model H~II region
grid of Sutherland, Shull, \& Beck (\markcite{s96}1996) to find the 
effective temperature of the exciting stars.  We use models with 
$z=0.1z_{\odot}$, the closest in the grid to the almost $0.2z_{\odot}$ 
oxygen value of He~2-10 (\markcite{c94}Conti \& Vacca 1994), and assume 
that Ne:Ar:S is solar.  The models have $log(n_e)=4.0$, which is typical 
of dense starburst regions, and log~U (the ionization parameter) 
-2.5 and -1.5.   

The [Ne~II]/[Ar~III] ratio is a good stellar temperature diagnostic for 
temperatures less than 45,000 K as it is fairly insensitive to density, 
ionization parameter and filling factor.  The [Ar~III] and [Ne~II] were 
not observed in identical locations nor at the same slit angles, so we 
have compared the total flux in the EW slit position on region A where 
[Ar~III] was observed to the total [Ne~II] flux observed in each of the 
NS [Ne~II] slit positions. This assumes that the distribution of the 
infrared lines is spherically symmetric, which within the resolution of
the observations it appears to be; the uncertainties introduced by possible 
small asymmetries will be less than those from other factors such as 
extinction.  The results uncorrected for reddening between $12.8$ and 
$8.99$ $\mu$m give stellar temperatures of 36-38,000 K for the different 
[Ne~II] slit positions. Any extinction correction will increase [Ar~III] 
relative to [Ne~II] and give hotter temperatures; both the Becklin et 
al.~(\markcite{b78}1978) and the Roche \& Aitken (\markcite{r84}1984) 
extinction curves give stellar temperatures 38-40,000 K. These results 
are very insensitive to the metallicity and will be the same within the 
uncertainties for solar metal content. The weakness of [Ne~II]/[Ar~III] 
ratio is that it is relatively insensitive to small temperature shifts 
in precisely the 36,000 K to 41,000 K range where He~2-10 falls.  The 
[Ne~II]/[S~IV] ratio (comparing the total observed [Ne~II] flux to the 
total observed [S~IV] flux) gives stellar temperatures above 37,000 K, 
but this ratio, although it has a large dynamic range, is very sensitive 
to the ionization parameter and the metallicity.  If we compare the 
ratios of [Ar~III], [S~IV], and [Ne~II] to those seen in the Galactic 
Center and NGC~3256 where the effective stellar temperature has been 
derived from the much more sensitive [Ne~II]/[Ne~III] ratio 
(\markcite{ku96}Kunze et al.~1996) and interpolate, we find a temperature 
of $39,000 \pm1400$ K, consistent with what the [Ne~II]/[Ar~III] ratio gives. 

The stellar temperatures (38,000-40,000 K) found from the infrared lines 
are equivalent to O6.5 ZAMS stars (\markcite{p73}Panagia 1973). The 
ionizing spectrum of a cluster of stars will differ from that of a single 
star, so it is only an approximation to treat the cluster as if it were 
composed only of one stellar type.  While we expect there to be many stars 
cooler than 40,000 K the weak [Ar~III] and [S~IV] rule out the presence 
of more than a very few stars hotter than 40,000. Cluster simulations 
(\markcite{ku96}Kunze et al.~1996) show that the upper mass cutoff in a 
cluster is typically about $5M_\odot$ greater than the single-star 
equivalent mass, so the upper mass cutoff is about $35M_\odot$. We can 
use Ho, Beck, \& Turner's (\markcite{h90}1990) result which relates 
the total luminosity of a star cluster to the ionization (with a slight 
correction because we use a different upper mass cutoff) and find that 
the total stellar luminosity needed for the Lyman continuum flux $N_{Lyc}$ 
of $7.6\times 10^{52}$ to $1.1\times10^{53}$ s$^{-1}$ (found from the 
extinction corrected $Br\alpha$ flux) is $3.7-5.3 \times 10^9L_\odot$.
This is quite close to the $8 \times 10^9L_\odot$ of total luminosity 
from the IRAS FIR flux of He~2-10.  The FIR beam of course contains the 
whole galaxy while the Brackett lines are limited to the core; He~2-10 is 
another example of the common starburst galaxy phenomenon that the bulk 
of the infrared luminosity observed by IRAS with large beams is generated 
by massive stars in a very small area (\markcite{h90}Ho et al.~1990).  

The stellar luminosity derived from the infrared lines may be compared 
to that observed at optical and ultraviolet wavelengths. The cluster 
observed in the infrared is about a factor of 10 more luminous than the 
5500 O stars that Vacca \& Conti (\markcite{v92}1992) find from optical 
spectra.  The ultraviolet observations of Conti \& Vacca (\markcite{c94}1994) 
find about 6 times as many O stars as their optical measurements; this 
may be due, as Conti \& Vacca suggest, to a combination of larger area 
coverage and contributions from evolved stars, but the dependance of 
their result on the very large extinction correction at 220nm may also 
be a factor; a small change in the proper extrapolation from $E_{B-V}$ 
to $A_{220}$ can change the final luminosity by more than a factor of 
two.  In any case, the luminosity of the star cluster derived from the 
infrared lines is at least twice as great and may be 5-10 times as great 
as the optical luminosity. This is the expected result when the extinction 
is so great that the optical measurements do not see through the source. 

The total mass of the obscured star cluster can be calculated from the
observed ionization if we assume a mass function. If we take a mass 
function that goes as $M^{-3.2}$, assume that stars smaller than $10_\odot$ 
do not contribute significantly to the ionization and sum the ionization of 
stars from 35 to 10 $M_\odot$ using Panagia's (\markcite{p73}1973) results 
for the ionizing flux of a ZAMS star of a given temperature and spectral 
type, we find the total mass of stars between $10M_\odot$ and $35M_\odot$ 
is $4.7-6.9\times 10^6M_\odot$ (where the spread in mass reflects the range 
of total ionization found from the Brackett lines). Giant and supergiant 
stars produce more ionization at a given spectral type; taking the extreme 
case that all the stars are luminosity class III reduces the total mass by 
about a factor of 2.  The ZAMS case implies a mass density of $8M_\odot pc^{-3}$ 
in a 130 pc radius, although since the young stars are probably clumped 
into small clusters like the ones seen in the UV (this may be confirmed 
when higher spatial resolution infrared observations become possible) 
this average density is only a lower limit.  Kobulnicky et 
al.~(\markcite{k95}1995) find a dynamical mass of $3.2\times10^6$ to 
$4.8\times10^7 M_\odot$ in the inner 70 pc, but caution that those are 
only lower limits due to beam size effects.  But even within the large 
uncertainties of both the CO and the infrared measurements, it is clear 
that a substantial fraction of the dynamical mass in the center is in 
the form of young massive stars.   
  
\subsection{Starburst History}

The effective stellar temperature of the cluster, as derived from the
mid-infrared lines above, can be used in combination with a set of
cluster evolution models to estimate the age and nature of the starburst.
The other important constraint on the starburst age is the non-thermal
radio flux, which we find by subtracting the 5 GHz thermal flux of 
$9-11 mJy$ (derived from the Brackett lines, the range reflects 
uncertainty in the extinction and the electron temperature) from the 
total $55mJy$  observed by Allen et al.~(\markcite{a76}1976). (While 
the radio fluxes were observed with much larger beams than the infrared, 
we will assume that the radio emission is dominated by the young star 
cluster and not by extended emission; this is usually the case in 
star-forming dwarfs.) We follow Allen et al. that the non-thermal 
radio emission is due to supernovae remnants rather than to the more 
extreme mechanisms which occur in radio galaxies; this agrees with 
the spectral index, the size and structure of the radio emission in 
the 3.6 cm maps, and the nature of the host galaxy. Then the 
non-thermal flux shows that a large number, which cannot be calculated 
exactly because of the wide range of supernova radio brightnesses but 
which we estimate at a few thousand, of supernovae have exploded in 
He~2-10. (We believe that the objection of Allen et al.~that a large 
number of supernovae is not consistent with the low metal content of 
the galaxy may not be valid; He~2-10 has a relatively high abundance 
of oxygen for a dwarf, 50\% higher than NGC~5253 for example, and the 
amount of O produced in the required number of supernova is consistent 
with the observed.  It would be highly desirable to observe the [O/Fe] 
ratio in He~2-10 to constrain the contribution of supernovae to the 
metal content.)   

The observations combined with the cluster evolution models of Sutherland
et al.~(\markcite{s96}1996) show that the cluster in He~2-10 is unlikely 
to have formed coevally.  A coeval star cluster cools off so fast that 
its effective temperature will be as high as the observed only at ages 
less than $2 \times 10^6$ yrs, which is not enough time for the SNe 
needed for the non-thermal radio emission. We turn to gaussian burst 
models, where the total mass of stars formed is a gaussian function of time, 
with its peak at $5 \times 10^6$yrs from the start of the burst and a 
FWHM in time of $4\times10^6$yrs, and the chances of a star of a given 
mass forming at a given time are found from Monte Carlo simulations.  
(Other scenarios for long-lived star bursts, such as exponentially 
decaying rates, are of course possible; the gaussian models are at 
present the best studied.) The effective stellar temperature found from 
the infrared lines does not constrain the burst age tightly; the cluster 
age can be from $5\times 10^6$ to $1 \times 10^7$ years since the start of 
the burst in the absence of WR stars and as old as $1.5 \times 10^7$ yrs 
if WR stars are important in the infrared cluster.  It similarly does not 
rule out longer-lived bursts.  (It is at present very difficult to detect 
WR stars in the infrared and there is only a very high upper limit on the 
possible number of WR stars in the infrared cluster (\markcite{l94}Lumsden, 
Puxley, \& Doherty 1994), nor can we assume that the stellar population in 
the obscured region is like that in the optically visible region.) If we 
assume that we do not see stars more massive than about $30M_\odot$ 
because they have become supernovae, the cluster must be at least 
$5 \times 10^6$ yrs old (\markcite{w93}Woosley, Langer \& Weaver 1993).  
The optical and infrared colors  indicate that the cluster is about 
$2 \times 10^7$ years old (Johansson \markcite{j87} 1987); this 
is somewhat older than our results for the infrared cluster but 
considering how ill-constrained the age of the infrared cluster is the 
discrepancy may not be significant. We should remember that only in very 
unusual cases can a star formation episode be given a unique age; the 
above numbers probably reflect a star formation rate generally high but 
fluctuating with time over the last $1-2\times10^7$ yrs. The optical 
cluster, for example, contains many WR stars much younger than the 
cluster age derived from the colors.  

\subsection{Conclusion}

We have found that the infrared emission of He~2-10 is produced by a 
star cluster $130\times180$ pc in size, with its ionizing radiation
dominated by O6.5 stars, and less than $1.5 \times 10^7$ years old.   
He~2-10 resembles NGC~5253, the most famous bright infrared WR galaxy, 
in that they are both dominated by condensed central clusters of 
young OB stars of similar (to within a factor of two) sizes.  It 
appears however to have had a very different history. The star cluster 
in NGC~5253 probably formed in a coeval burst; star formation in He~2-10 
appears to have been a more sustained process. The galaxies are also at
different evolutionary stages; the star cluster in He~2-10 may be almost 
10 times as old as that in NGC~5253. The young star cluster in He~2-10 
has a disk shape and lies at the same angle as the steepest velocity 
gradient of the recently accreted molecular gas, which is different 
from the orientations of the galactic halo and the bulk of the CO. 
We believe that the recent burst of star formation we have observed 
lies in the original galactic disk and was triggered by the infall of 
molecular gas from  the accretion event. The offset between the disk 
and the halo may be the result of the merger process or may have 
predated the merger.

\acknowledgements 

We thank Mr.~Charles Kaminski for obtaining the infrared images and 
the NASA IRTF for the time, Dr.~J.M.~van der Hulst for the use of the 
[S~IV] data, and Mr.~H.A.~Kobulnicky for access to unpublished data 
and for interesting discussions.  This work was supported by the 
US-Israel Binational Science Foundation grant 94-00303 and by USAF 
contract F19628-93-K-0011. D.K. is supported at Wyoming by NSF grant 
AST94-53354.

\newpage

\figcaption{The spectrum of [Ne~II] observed with a 
2$^{\prime\prime} \times$ 11$^{\prime\prime}$ NS slit across the 
center of peak~A.  \label{fig1}}

\figcaption{J, H, and K band images of He~2-10 obtained with NSFCAM on 
the IRTF.  a) J band: the first contour is at 16000 counts and the 
interval is 2500 counts.  b) H band: the first contour is at 53000 
and the interval is 2000.  c) K band: the first contour is at 55000 
and the interval is 3000.  
\label{fig2}}

\figcaption{The logarithms of the a)J/H, b)J/K, and c)K/H map ratios.
These are equivalent to the relative J-H, J-K and K-H colors up to 
a constant. \label{fig3}}

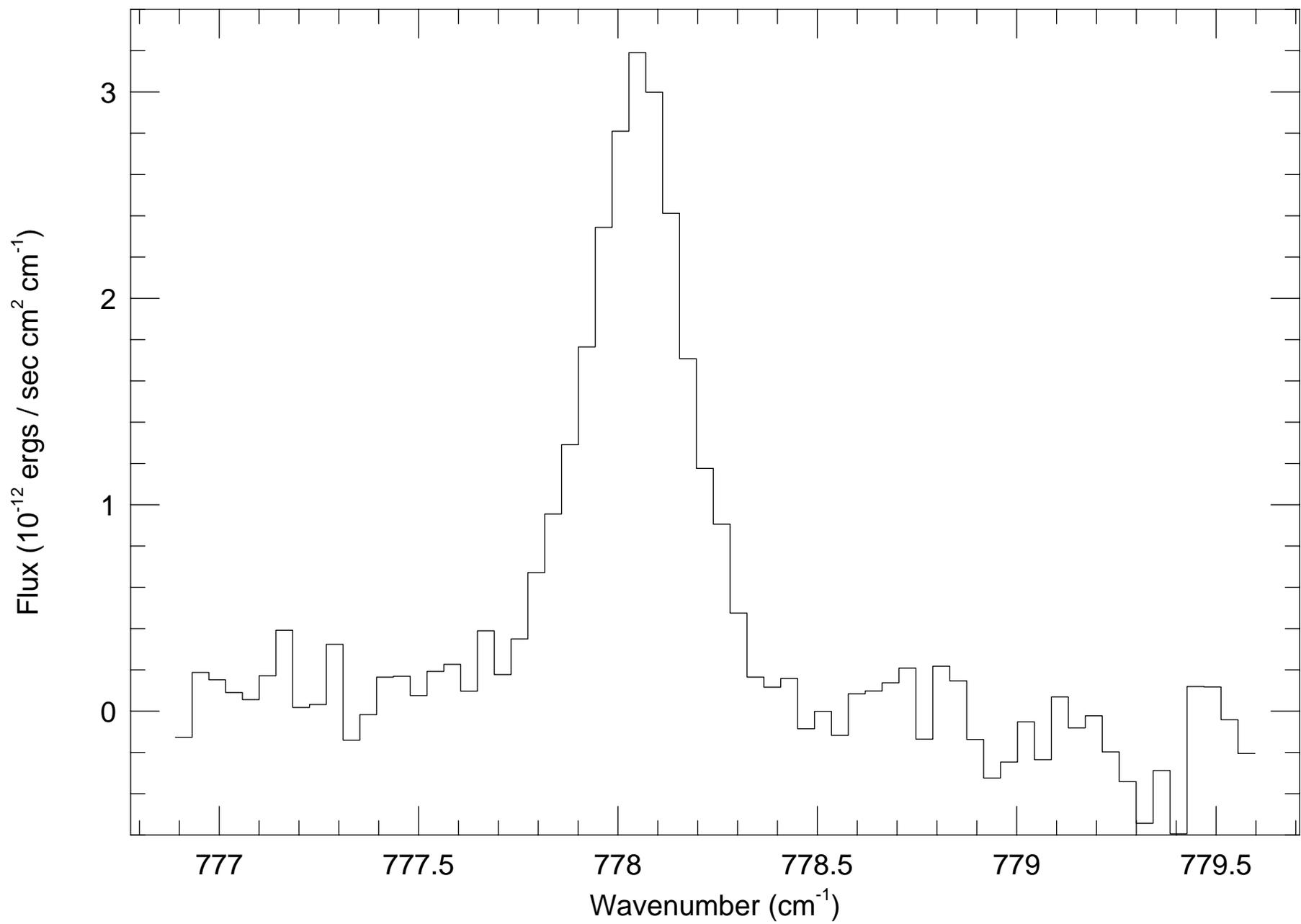
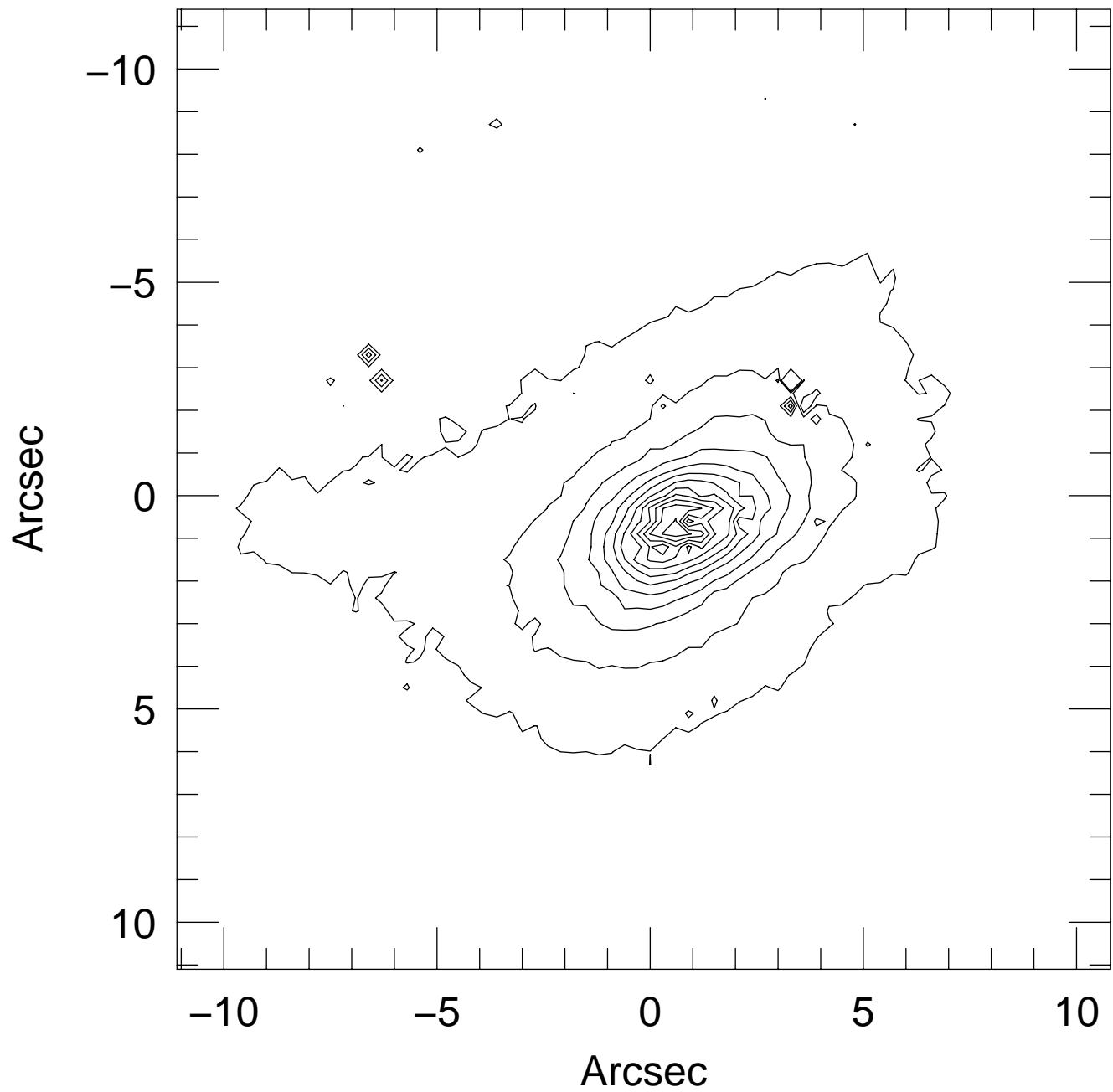
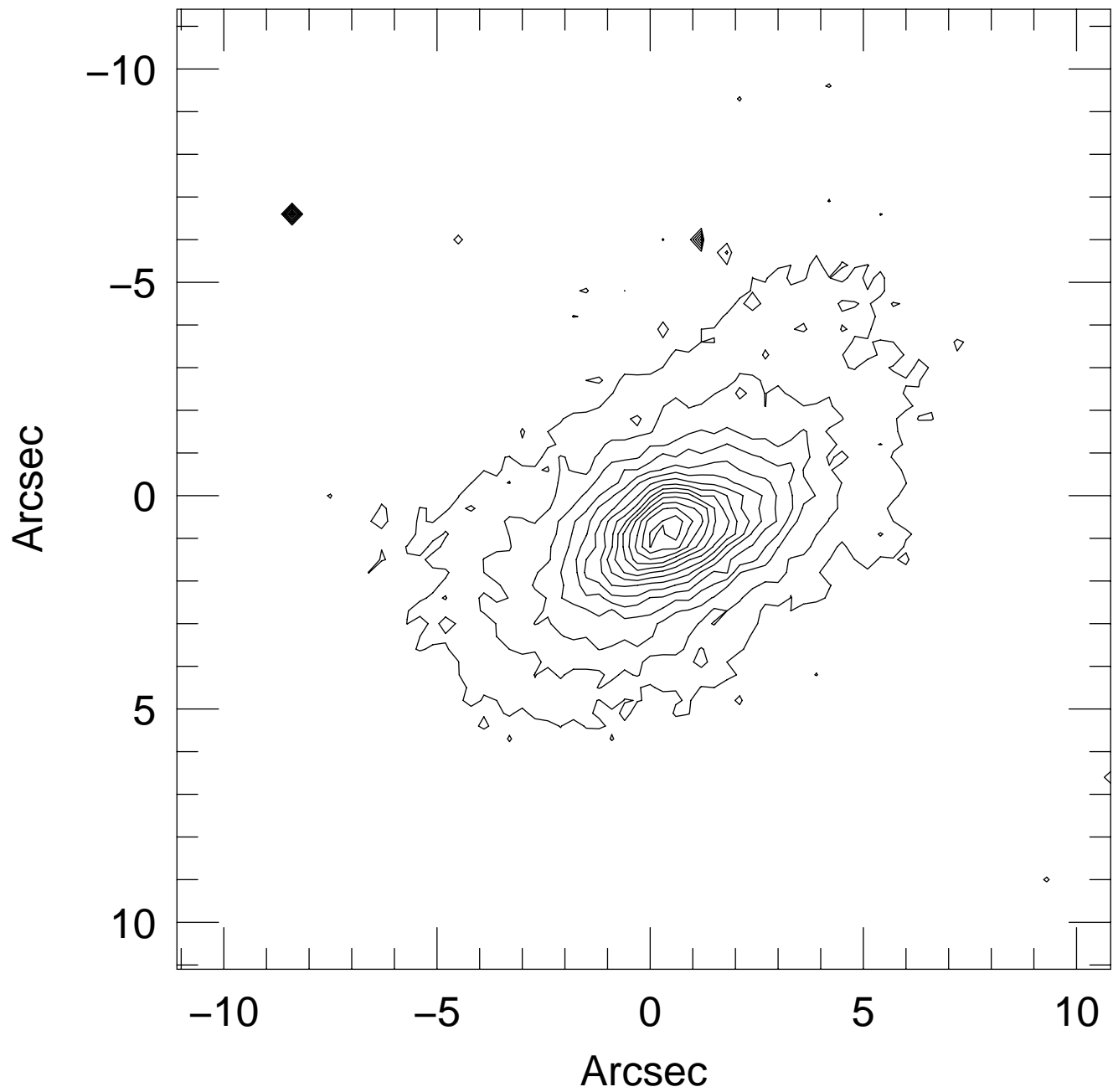
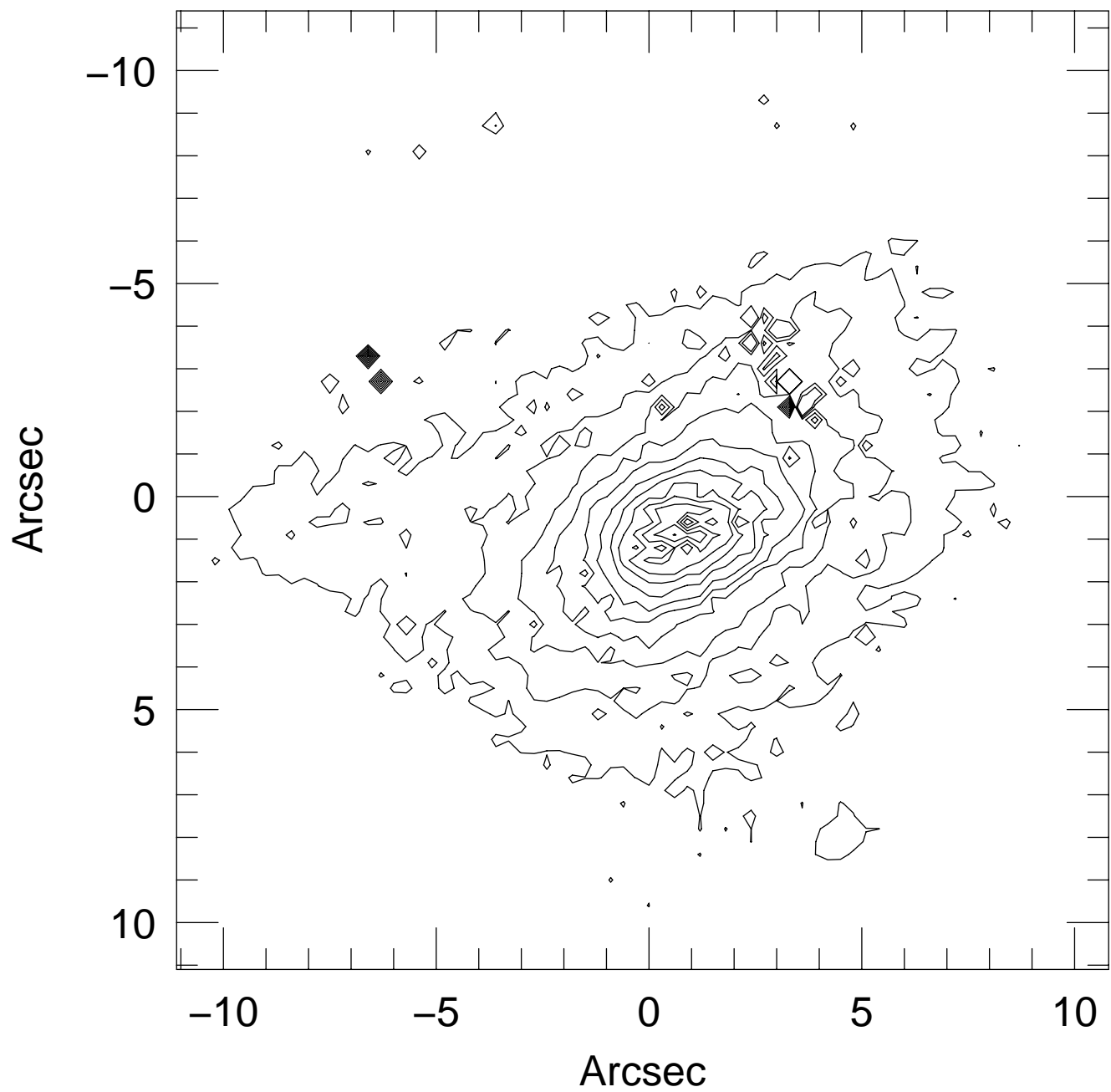
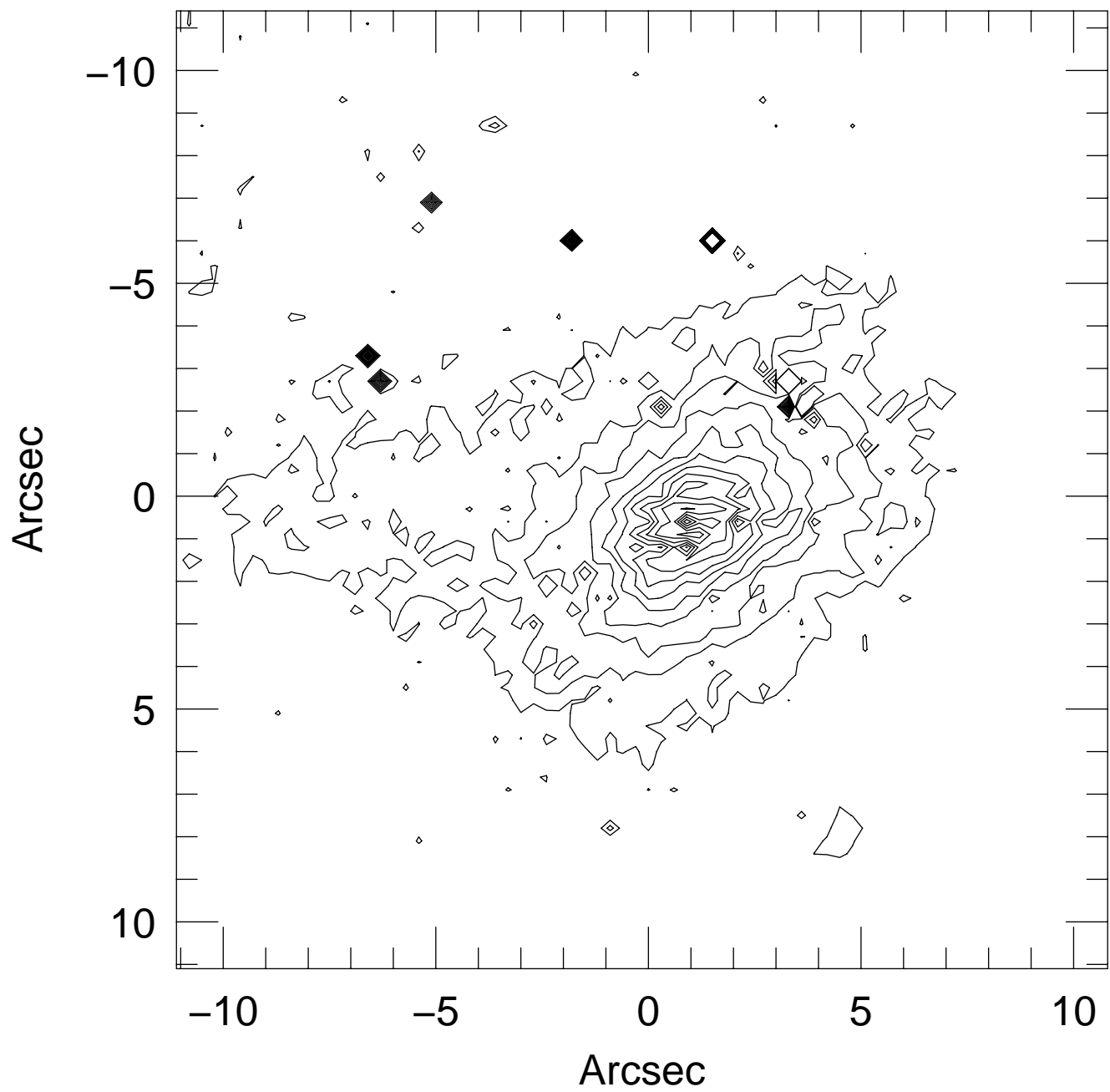
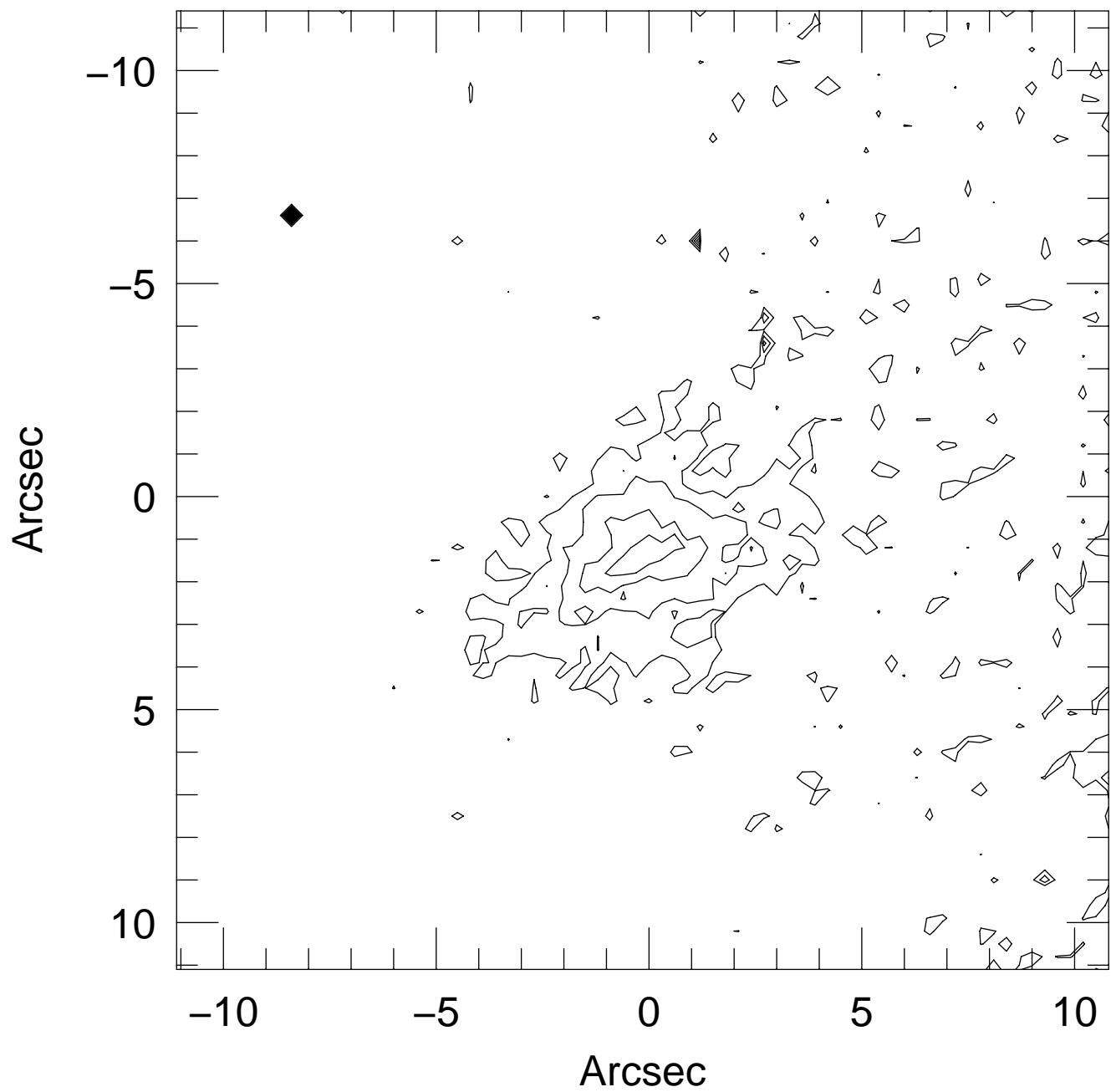
\end{document}